\documentstyle[twocolumn,prl,aps,floats]{revtex}
\input psfig

\def\bea{\begin{eqnarray}}
\def\eea{\end{eqnarray}}
\def\ben{\begin{equation}}
\def\een{\end{equation}}
\def\benu{\begin{enumerate}}
\def\enu{\end{enumerate}}

\def\n{n}

\def\sss{\scriptscriptstyle\rm}

\def\1var{(\bx_1...\bx\N)}

\def\half{\frac{1}{2}}

\def\br{{\bf r}}

\def\bx{{x}}

\def\x{_{\sss X}}
\def\c{_{\sss C}}
\def\s{_{\sss S}}
\def\xc{_{\sss XC}}
\def\Hxc{_{\sss HXC}}

\def\N{_{\sss N}}

\def\LDA{^{\rm LDA}}

\def\unif{^{\rm unif}}

\def\up{_\uparrow}
\def\dn{_\downarrow}
\def\up{_\alpha}
\def\dn{_\beta}

\def\sph_int{ {\int d^3 r}}

\def\JCP{J. Chem. Phys.\ }

\input rotate

\def\up{\uparrow}
\def\dn{\downarrow}
\def\Hc{_{\sss HC}}
\def\n{\rho}

\begin{document}

\twocolumn[\hsize\textwidth\columnwidth\hsize\csname
@twocolumnfalse\endcsname
\title{\bf A hybrid functional for the exchange-correlation kernel
in time-dependent density functional theory}
\author{Kieron Burke}
\address{Departments of Chemistry and Physics, Rutgers University, 610
Taylor Road, Piscataway, NJ 08854}
\author{M.~Petersilka, and E.~K.~U.~Gross}
\address{Institut f\"ur Theoretische Physik, Universit\"at W\"urzburg,
Am Hubland, 97074 W\"urzburg, Germany}
\date{In preparation for DFT99 proceedings, \today}
\maketitle
\begin{abstract}
A review of the approximations in any time-depedendent
density functional calculation of excitation energies is
given.   The single-pole approximation for the susceptibility
is used to understand errors in popular approximations for
the exchange-correlation kernel.  A new hybrid of exact
exchange and adiabatic local density approximation is proposed
and tested on the He and Be atoms.
\end{abstract}
]
\narrowtext

\section{Introduction}
\label{intro}

Ground-state density functional theory is well-established
as an inexpensive alternative to traditional {\em ab initio}
quantum chemical methods\cite{KP98}.  Now
time-dependent density functional theory (TDDFT) is
rapidly emerging as an inexpensive accurate method
for the calculation of electronic excitation energies
in quantum chemistry\cite{GDP96,BG98}.
Calculation of dynamic response properties using TDDFT has
a long history, since the pioneering work of Zangwill and
Soven\cite{ZS80,E84}.  It is only relatively recently that
attention has been focussed on the direct extraction of excitation
energies\cite{PGG96,C96,PG96,G97,JCS96,GPG00}.  Already, this method
has been implemented in several quantum chemistry packages, such as
deMon\cite{CCS98}, Turbomole
\cite{BA96,BHTA97},
ADF\cite{GKSG98},
and QCHEM\cite{HH99}.
Important calculations include the calculation of
excited-state crossings in formaldehyde\cite{CCS98},
excitations with significant doubly-excited character\cite{HH99},
the photospectrum of
chlorophyll A\cite{Sb99}, and the response of 2-D quantum
strips\cite{UV98}.

How are excitation energies calculated using TDDFT?
First, a self-consistent ground-state Kohn-Sham calculation is performed,
using some approximation for the exchange-correlation energy $E\xc$,
such as B3LYP\cite{B96} or 
PBE\cite{PBE96}.  This yields a set of Kohn-Sham eigenvalues $\epsilon_i$
and orbitals $\phi_i$.  Even with the {\em exact} ground-state energy
functional and potential, these eigenvalues are in general {\em not} the
true excitations of the system, but are closely related.
In a second step, 
the central equation of TDDFT response theory
is solved, which extracts the true linear response function
from its Kohn-Sham counterpart.
This equation includes a second unknown functional, the
exchange-correlation
kernel $f\xc(\br,\br';\omega)$, which is the Fourier transform of the functional
derivative of the {\em time-dependent} exchange-correlation potential.
The poles of the exact response function are shifted from
those of the KS function, and occur at the true excitations of
the system.
These steps are typically repeated for several nuclear positions.

The success of any density functional method, however, depends
on the quality of the approximate functionals employed.               
The above calculation requires two distinct density functional
approximations:  one for the ground-state energy, which implies a
corresponding approximation for the exchange-correlation
potential $v\xc(\br)=\delta E\xc/\delta \n(\br)$, and a second for the
exchange-correlation kernel.
Most calculations now appearing in the chemical literature
use the adiabatic local density approximation (ALDA) for $f\xc$.
Adiabatic implies that the frequency-dependence of $f\xc$
is ignored, and its $\omega=0$ value used, and LDA implies
$f\xc(\br,\br')=\delta v\xc(\br)/\delta \n(\br')$,
where $v\xc$ is the LDA potential from ground-state LDA calculations.

The more drastic of the two approximations is that for
the ground-state.  Very often, functionals which yield accurate
ground-state energies have very poor-looking potentials.  (How this can
happen can be understood by considering the virial theorem, which
relates energies to potentials\cite{BCL98}).  These
poor potentials then have in turn badly
behaved virtual orbitals.  In particular, local and semi-local
approximations, i.e., generalized gradient approximations, fail to
capture the correct asymptotic behavior of the potential, and many
virtual states are not even bound.  Even hybrid functionals, such
as B3LYP\cite{B96} and 
PBE0\cite{AB99}, do not much improve the
asymptotic behavior, since they only mix a fraction of exact exchange.
This restricts calculations within
these approximations to only low-lying excitations.  This difficulty is
most pronounced in atoms, becomes smaller for bigger molecules, and  is
irrelevant for bulk solids.  Recently\cite{PGB00}, we have shown that even with
approximations which are free from self-interaction error, such as
exact exchange, or self-interaction corrected LDA, and
which therefore reproduce the dominant
part of the asymptotic decay of the potential, inaccuracies in the
Kohn-Sham eigenvalues (mostly due to the incorrect position of the
highest occupied level) dominate over any errors introduced in the second
step, namely the correction of KS levels to the true levels.

The present work studies the accuracy of approximations to $f\xc$
alone.  This is because, for low-lying states, errors in $f\xc$ must
be disentangled from errors in the KS energy levels, and also because
many people are working to improve approximations to
the exact ground-state potential\cite{CJCS98,IH96,TH98},
which we hope will ultimately reduce
those errors discussed above.   Hence all our calculations are performed using
the {\em exact} Kohn-Sham potentials of the He and Be atoms, for which we thank Cyrus
Umrigar\cite{UG94}.  We study a variety of approximations to $f\xc$, all adiabatic.
We focus especially on the spin-decomposition of such approximations,
which determines the relative positions of singlet and triplet
excitations in TDDFT.   We use the single-pole approximation for the
susceptibility to directly relate errors in $f\xc$ to errors in 
excitation energies.  We find that exact exchange works well for
parallel spins, which determines the mean energy of the singlet and
triplet, while the antiparallel contribution to $f\xc$ determines
their splitting, which is well-approximated in ALDA.
With this insight, we construct a hybrid of exact
exchange and ALDA, which greatly improves results for He, and moderately
improves them for Be.  We use atomic units ($e^2=\hbar=m_e=1$)
throughout, except in Fig. 1.

\section{Methodology}

The basic response equation of TDDFT has the same form
as that of time-dependent Hartree-Fock theory, or of the
Random Phase Approximation, i.e.,
\bea
\chi^{\sigma\sigma'}(\br,\br';\omega)&=&\chi\s^{\sigma\sigma'}(\br,\br';\omega)
\nonumber\\
&+&\sum_{\sigma''\sigma'''}
\int d^3r''\ \int d^3r'''\ \chi\s^{\sigma\sigma''}(\br,\br'';\omega)\
\nonumber\\
&& f\Hxc^{\sigma''\sigma'''}(\br'',\br''';\omega) \
\chi^{\sigma'''\sigma'}(\br''',\br';\omega)
\label{RPA}
\eea
where $\chi$ is the exact frequency-dependent susceptibility of the
system, while $\chi\s$ is its Kohn-Sham analog, and
$f\Hxc^{\sigma\sigma'}(\br,\br')=1/|\br-\br'|+f\xc^{\sigma\sigma'}(\br,\br')$.
The $\sigma$ indices denote spin, i.e., $\sigma=\uparrow$ or $\downarrow$.
By various means,
the poles of $\chi$ as a function of $\omega$ can be found\cite{C96}.
Our method for finding these poles is to
consider only the discrete poles, i.e., those corresponding to
bound states.  In the particular case of a frequency-independent model
for $f\xc$, we can show\cite{PGB00} that these poles occur at the
eigenvalues of the matrix 
\bea
\label{M}
M^{\sigma\sigma'}_{qq'}&=&\delta_{qq'}^{\sigma\sigma'}\omega_{q\sigma}+
\alpha_{q' \sigma'}
\int\!d^3 r \int\!d^3 r' \,
\Phi^{*}_{q \sigma}({\bf r})\times
\nonumber\\
&&
f\Hxc^{\sigma \sigma'}({\bf r},{\bf r'})
\Phi_{q' \sigma'}({\bf r'})  \,.
\eea
For notational brevity, we have used double indices $q\equiv(j,k)$ to
characterize the excitation energy
$\omega_{q \sigma}\equiv\epsilon_{j \sigma}-\epsilon_{k
\sigma}$ of the single-particle
transition $(k \sigma \rightarrow j \sigma)$.
Consequently, we set
$ \alpha_{q \sigma} := f_{k \sigma}-f_{j \sigma} $,
where $f_{k \sigma}$ is the occupation number of that orbital,
and
$\Phi_{q \sigma}({\bf r})
=
\varphi_{k \sigma}^{*}({\bf r})\varphi_{j \sigma}({\bf r})$.
While we do include sufficient bound-state poles to converge
to an accurate result, our method does neglect continuum contributions,
and this effect will be discussed in the next section.

All approximations we study for $f\xc$ are adiabatic.
The most ubiquitous is ALDA (or more precisley, the adiabatic
local spin density approximation)
in which
\ben
f\xc^{\sigma \sigma' \rm ALDA}({\bf r},{\bf r'})
=\delta (\br-\br')\ 
\frac{\partial^2 e\xc\unif(n_\up,n_\dn)}{\partial n_\sigma
\partial n_{\sigma'}} \Big|_{n_\up(\br),n_\dn(\br')}.
\een
Note that this leads to a completely short-ranged approximation
to $f\xc$.  Similarly, any adiabatic GGA approximation leads
to an approximate $f\xc$ which is almost as short-ranged.

A second distinct approximation to $f\xc$ is in terms
of its (usually) dominant exchange contribution.    A highly
accurate approximation to the exact exchange-only equations
of ground-state density functional theory
(the optimized effective potential
equations) was introduced by Krieger, Li, and Iafrate\cite{KLI90}.
This approximation
has been extended to the time-dependent case\cite{PGG96}:
\begin{equation}
\label{fxcoep}
f\x^{\sigma\sigma'}({\br},\br')=
- \delta_{\sigma\!\sigma'}
{\big|\sum_k
f_{k\sigma}\,\varphi_{k\sigma}({\br})\varphi_{k\sigma}^*({\br'})\big|^2
\over n_\sigma({\br}) |{\br}-{\br'}|n_\sigma({\br'})}\;.
\end{equation}
This is exact for one electron, and for (spin-unpolarized)
two-electron exchange.
Note that this approximation has
a long-ranged contribution, which can cancel exactly the direct hartree
contribution to the matrix ${\bf M}$ in Eq. (\ref{M}).

A third approximation which we tried is the self-interaction
corrected (SIC) ALDA, which is simply the second functional derivative
of the SIC-LDA energy:
\ben
E\xc^{\rm SIC-LDA} = E\xc\LDA + \sum_{i\sigma} (E\x[n_{i\sigma}]-
E\xc\LDA[n_{i\sigma}]),
\een
where $n_{i\sigma}=|\varphi_{i\sigma}|^2$ is the density of a single-orbital,
and $E\x[n_{i\sigma}]$ is the exact Hartree self-interaction energy of
that orbital.  This approximation should improve over ALDA, in avoiding
spurious self-interaction errors, and over just exchange, by including
some correlation.

\section{Data}

%
\begin{table}[hbt]
\caption{\label{tabext}
Singlet/triplet excitation energies in the helium and beryllium atoms,
calculated from the exact Kohn-Sham potential by
using approximate xc kernels (in millihartrees), and
using the lowest 34 unoccupied orbitals of s and p symmetry for He,
and the lowest 38 unoccupied orbitals of s, p, and d symmetry for Be.
Exact values from Ref. \protect
\cite{KH84} for He and from Ref. \protect
\cite{BS75} for Be.}
\begin{tabular}{ccccccc}
&&\multicolumn{5}{c}{\rm Singlet/triplet shifts}\\
\hline 
&{$\omega_{KS}$}
      &\multicolumn{1}{c}{\rm ALDA} 
      &\multicolumn{1}{c}{\rm X} 
      &\multicolumn{1}{c}{\rm SIC}
      &\multicolumn{1}{c}{\rm hybrid}
      &\multicolumn{1}{c}{\rm exact}\\
\hline
\multicolumn{7}{c}{\rm Transitions from the $1s$ state in He atom}\\
\hline
{$2s$}& 746.0& 22/-11& 20/-25& 19/-16& 14/-19& 12/-18\\
{$3s$}& 839.2&  6.9/-2.4&  5.8/-4.9&  5.6/-3.6&  5.1/-4.6&  3.3/-4.2\\
{$4s$}& 868.8&  3.1/-0.9&  2.5/-1.7&  2.4/-1.3&  2.4/-2.0&  1.3/-1.6\\
{$5s$}& 881.9&  1.6/-0.4&  1.3/-0.8&  1.3/-0.6&  1.3/-1.0&  0.6/-0.8\\
{$6s$}& 888.8&  1.0/-0.3&  0.8/-0.5&  0.7/-0.4&  0.8/-0.7&  0.4/-0.5\\
\hline
{$2p$}& 777.2& -0.8/-7.4&  7.2/-8.4&  6.1/ 0.2&  2.7/-3.4&  2.7/-6.6\\
{$3p$}& 847.6&  0.7/-1.9&  2.5/-2.3&  2.2/-0.5&  1.4/-1.3&  1.0/-2.0\\
{$4p$}& 872.2&  0.4/-0.7&  1.1/-0.9&  1.0/-0.2&  0.6/-0.5&  0.5/-0.8\\
{$5p$}& 883.6&  0.2/-0.4&  0.6/-0.5&  0.5/-0.1&  0.4/-0.3&  0.2/-0.4\\
{$6p$}& 889.8&  0.1/-0.3&  0.3/-0.3&  0.3/-0.1&  0.2/-0.2&  0.1/-0.3\\
\hline
err &57& 32 & 31&  29 & 12& -\\
\hline
\multicolumn{7}{c}{\rm Transitions from the $2s$ state in Be atom}\\
\hline
{$3s$}&244.4&  7.1/-5.7& 10.9/-10.6& 10.3/-1.4&  6.6/-4.6&  4.7/-7.1\\
{$4s$}&295.9&  2.5/-1.6&  3.6/-2.5&  3.5/-0.6&  2.6/-1.6&  1.4/-2.0\\
{$5s$}&315.3&  1.1/-0.7&  1.7/-1.0&  1.6/-0.3&  1.2/-0.7&  0.6/-0.9\\
{$6s$}&324.7&  0.6/-0.4&  0.9/-0.5&  0.9/-0.2&  0.7/-0.4&  0.3/-0.5\\
\hline
{$2p$}&132.7& 56/-42& 55/-133& 53/-53& 10/-88& 61/-32\\
{$3p$}&269.4&  2.0/-4.3&  6.4/-4.2&  5.6/ 1.1&  4.2/-2.8&  4.8/-1.5\\
{$4p$}&304.6&  0.3/-1.4&  2.1/-1.2&  1.9/ 0.3&  1.3/-0.8&  1.7/-4.1\\
{$5p$}&319.3&  0.1/-0.6&  1.0/-0.5&  0.9/ 0.1&  0.6/-0.2&  0.2/ 0.0\\
{$6p$}&326.9&  0.0/-0.4&  0.5/-0.3&  0.4/ 0.0&  0.3/-0.2&  0.1/-0.1\\
\hline
{$3d$}&283.3& -5.4/-2.8&  1.8/-2.0&  0.9/ 3.2& -1.4/ 1.2& 10.3/-0.6\\
{$4d$}&309.8& -1.4/-1.1&  0.8/-0.9&  0.5/ 0.9& -0.2/ 0.6&  3.6/-0.2\\
\hline
err& 138&56 &144 &73&136 & - \\
\hline
err'& 45&41  &37 &44  &29& - \\
\end{tabular}
\end{table}
In this section, we report calculations for the He and Be atoms
using the exact ground-state Kohn-Sham potentials,
the three approximations to the kernel
mentioned in the previous section, and
including many bound-state poles in Eq. (\ref{M}),
but neglecting the continuum.
The technical details are given
in Ref. \cite{PGB00}.
Table I lists the results, which
are compared with a highly accurate
nonrelativistic variational calculations\cite{KH84,BS75}
In each symmetry class (s, p, and d), up to 38 virtual states were
calculated.  The errors reported are absolute deviations from the
exact values.  The second error under the Be atom excludes
the $2s\to 2p$ transition, for reasons discussed in the next
section.

The effect of neglecting continuum states in these calculations
has been investigated by van Gisbergen et al.\cite{GKSG98},
who performed ALDA calculations from the exact Kohn-Sham potential
in a localized basis set.  These calculations were done including
first only bound states, yielding results identical to those
presented here, and then including all positive energy orbitals
allowed by their basis set.  They found significant improvement
in He singlet-singlet excitations, especially for $1s\to 2s$ and
$1s\to 3s$.   Other excitations barely changed.   Assuming inclusion
of the continuum affects results with other approximate kernels similarly,
these results do not change the basic reasoning and conclusions
presented below, but suggest that calculations including the continuum
may prove to be more accurate than those presented here.

\section{Single-pole analysis}

The simplest truncation of the eigenvalue equation (\ref{M})
for the excitation energies is to ignore all coupling
between poles, except that between a
singlet-triplet pair.  This is equivalent to
setting $\langle q | f\Hxc | q' \rangle$ to zero,
for $q\neq q'$.  (We have dropped the spin-index on these
contributions, since we deal only with closed shell systems).
Then the eigenvalue problem reduces to a simple $2\times 2$
problem, with solutions
\bea
\Omega_q^+ &=& \omega_q + 2 \Re \langle q | f\Hxc | q \rangle
\nonumber\\
\Omega_q^- &=& \omega_q + 2 \Re \langle q | \Delta f\xc | q \rangle,
\eea
where 
\bea
f\Hxc&=&\frac{1}{4}\sum_{\sigma\sigma'} f\Hxc^{\sigma\sigma'}=
\frac{1}{|\br-\br'|}+\half \left(f\xc^{\uparrow\uparrow}+f\xc^{\downarrow\uparrow}\right),
\nonumber\\
\Delta f\xc&=&\frac{1}{4}\sum_{\sigma\sigma'} 
\sigma\sigma'f\Hxc^{\sigma\sigma'}=
\half \left(f\xc^{\uparrow\uparrow}-f\xc^{\downarrow\uparrow}\right).
\eea
Thus $f\Hxc$ is the spin-summed contribution, which contributes to
$\chi$, the spin-summed susceptibility, and therefore gives rise to
the singlet level, while $\Delta f\xc$
is the spin-flip contribution, also called
$\mu_o^2 G_{\rm xc}$ in the theory of the 
frequency-dependent magnetization density 
\cite{LV89}.
Thus even within the SPA, the KS degeneracy  between singlets 
and triplets is broken, and we identify $\Omega_q^-$ with the
triplet.
In Table II, we report results within the single-pole
approximation.
\begin{table}[hbt]
\caption{\label{tabSPA}
Same as Table I, but within the single pole approximation.}
\begin{tabular}{cccccc}
\hline
&\multicolumn{1}{c}{$\omega_{KS}$}&\multicolumn{4}{c}{\rm Singlet/triplet shifts}\\
\hline 
&exact
      &\multicolumn{1}{c}{\rm ALDA} 
      &\multicolumn{1}{c}{\rm X} 
      &\multicolumn{1}{c}{\rm SIC}
      &\multicolumn{1}{c}{\rm hybrid}\\
\hline
\multicolumn{6}{c}{\rm Transitions from the $1s$ state in He atom}\\
\hline
{$2s$}& 746.0	& 25.8/-10.3& 22.7/-22.8& 21.4/-14.7& 18.0/-18.1\\
{$3s$}& 839.2	&  6.6/-2.6&  5.6/-5.5&  5.3/-3.9&  4.7/-4.5\\
{$4s$}& 868.8	&  2.6/-1.0&  2.2/-2.1&  2.1/-1.5&  1.8/-1.8\\
{$5s$}& 881.9	&  1.3/-0.5&  1.1/-1.1&  1.0/-0.8&  0.9/-0.9\\
{$6s$}& 888.8	&  0.7/-0.3&  0.6/-0.6&  0.6/-0.4&  0.5/-0.5\\
\hline
{$2p$}& 777.2	& -0.8/-7.0&  7.8/-7.9&  6.4/ 0.2&  3.1/-3.1\\
{$3p$}& 847.6	&  0.7/-2.0&  2.4/-2.3&  2.1/-0.5&  1.4/-1.3\\
{$4p$}& 872.2	&  0.4/-0.8&  1.0/-1.0&  0.9/-0.2&  0.6/-0.6\\
{$5p$}& 883.6	&  0.2/-0.4&  0.5/-0.5&  0.5/-0.2&  0.3/-0.3\\
{$6p$}& 889.8	&  0.1/-0.3&  0.3/-0.3&  0.2/-0.1&  0.2/-0.2\\
\hline
err &57& 34 & 32& 31 & 15\\
\hline
\multicolumn{6}{c}{\rm Transitions from the $2s$ state in Be atom}\\
\hline
{$3s$}&244.4	&  8.2/-5.4& 13.0/-9.5& 12.2/-1.3&  8.6/-5.1\\
{$4s$}&295.9	&  2.4/-1.6&  3.5/-2.7&  3.4/-0.6&  2.4/-1.6\\
{$5s$}&315.3	&  1.0/-0.7&  1.5/-1.1&  1.4/-0.3&  1.1/-0.7\\
{$6s$}&324.7	&  0.5/-0.4&  0.7/-0.6&  0.7/-0.2&  0.5/-0.4\\
\hline
{$2p$}&132.7	& 75/-34& 71/-65& 67/-41& 58/-52\\
{$3p$}&269.4	& -0.4/-4.3&  5.4/-4.7&  4.5/ 0.6&  2.3/-1.6\\
{$4p$}&304.6	& -0.1/-1.5&  1.7/-1.5&  1.5/ 0.2&  0.8/-0.6\\
{$5p$}&319.3	& -0.1/-0.7&  0.8/-0.6&  0.7/ 0.1&  0.4/-0.2\\
{$6p$}&326.9	& -0.1/-0.4&  0.4/-0.4&  0.3/ 0.0&  0.1/-0.2\\
\hline
{$3d$}&283.3	& -5.0/-2.6&  1.8/-1.9&  1.0/ 3.4& -1.2/ 1.1\\
{$4d$}&309.8	& -1.4/-1.1&  0.8/ 0.0&  0.6/ 0.8&  0.3/ 0.6\\
\hline
err& 138&61 &79 &60&55 \\
err' & 45&45  &36 &44  &33\\
\end{tabular}
\end{table}

At this point, we notice the very strong shift in the Be $2s\to 2p$ transition.
This is due to the small magnitude of its transition energy,
so that the pole of the $2p$ energy is very close to the pole of the $2s$
energy.   Thus the single pole approximation is not expected to work
well for this case, and it should be excluded from general statements based
on the SPA.

\subsection{Why are Kohn-Sham excitation energies so good?}

We see throughout the data that the Kohn-Sham eigenvalues are always inbetween
the exact singlet and triplet
energy levels.  The splitting is much larger for Be than for He, but
this obervation is true in both cases.  It has already been made by Filippi
et al\cite{FUG97}, and explained in terms
of quasi-particle amplitudes\cite{SUG98}.
Here, we use the single-pole approximation to analyze
this result in terms of the known behavior of density functionals.
From Eq. (6) we find that the mean energy is given by
\ben
\bar\Omega_q=\half (\Omega_q^+ + \Omega_q^- )=\omega_q+\Re 
\langle q |
\frac{1}{|{\bf r}-{\bf r'}|}
+ f\xc^{\uparrow\uparrow}({\bf r},{\bf r'}
| q \rangle,
\label{mean}
\een
while the energy splitting is given by
\ben
\Delta\Omega_q=(\Omega_q^+ - \Omega_q^-)=2\Re
\langle q | \frac{1}{|\br-\br'|}+ f\c^{\uparrow\downarrow}| q \rangle,
\label{split}
\een
since there is no exchange contribution to antiparallel $f\xc$.
If we further define $\delta \Omega_q = \bar\Omega_q-\omega_q$
as the deviation of the mean energy from the Kohn-Sham level,
we see that 
\ben
|\delta \Omega_q| < \Delta \Omega_q/2
\label{inbetween}
\een
must be satisfied for the Kohn-Sham level to lie in between the
singlet and triplet levels.  Within the single-pole approximation,
we have a very simple expression for the ratio of these two:
\ben
\frac {2\delta\Omega_q}{\Delta \Omega_q} = 
\frac{ \Re \langle q | f\Hxc^{\uparrow\uparrow} | q \rangle}
{ \Re \langle q |
f\Hc^{\uparrow\downarrow} | q \rangle}
\label{ratio}
\een

Consider first the He atom.  For two electrons, $f\x^{\uparrow\uparrow}
= - 1/|\br-\br'|$, exactly cancelling the Hartree term in Eq. (\ref{mean}),
leaving only the parallel correlation contribution.
(This is reflected in the X column on Table II, where the upshift
of the singlet is equal to the downshift of the triplet.)
Thus we find
\ben
\frac {2\delta\Omega_q}{\Delta \Omega_q} = 
\frac{ \Re \langle q | f\c^{\uparrow\uparrow} | q \rangle}
{ \Re \langle q |
\frac{1}{|{\bf r}-{\bf r'}|}+
f\c^{\uparrow\downarrow} | q \rangle}
~~~~~~~~~{\rm (2 el)}
\label{tworatio}
\een
for two electrons in the single-pole approximation.  It is well-known
(see, e.g., Ref. \cite{WPb91}) that for ground-state energies, 
parallel correlation is much weaker than antiparallel, since
antiparallel electrons are not kept apart by the exchange interaction.
Thus this ratio is expected to stay well less than 1, as it does for all
our He excitations.  The effect of the single-pole approximation on this
conclusion can be judged by studying the shift in the mean for the X
results in Table I.

For Be, and any system with more than two electrons, there is still a
good deal of cancellation of the exchange contribution with the
direct contribution, but this cancellation is no longer exact.  This
can be seen in the Be results for X in Table II.  By studying the form
of $f\x$ given in Eq. (4), we expect this remnant exchange contribution
to be of order $O( (N-2)/N^2)$ for unpolarized systems.  Thus in
exchange-dominated (i.e., high density or weakly correlated) systems,
the direct Coulomb term in the denominator will be larger than
any remnant exchange term in the numerator.  On the other hand,
in low-density or strongly correlated systems, if antiparallel
correlation continues to dominate over parallel correlation,
this ratio will still be less than one.  We conclude that the
Kohn-Sham levels will usually be close to the true excitations
(of single-particle nature).

\subsection{Relation of exact exchange to G{\"o}rling-Levy perturbation theory}

Both time-dependent DFT and G{\"o}rling-Levy
\cite{GL93} perturbation theory are
formally exact methods for extracting electronic excitation energies
in density functional theory.  In this section, we consider the expansion
of the excitation energies in powers of the adiabatic coupling constant
$\lambda$ to first order.  This procedure should give identical
results in both theories.  Recently, Filippi et al\cite{FUG97}
have performed GL first
order calculations for the He atom, using the exact Kohn-Sham potential.
Their results are numerically identical to ours,
using the exact exchange kernel,
but only {\em within} the single-pole approximation, as given by Table II.
Results calculated with the full (i.e., many poles) scheme, i.e., in
Table I, differ slightly from
theirs.  This produces a paradox, in which the easier SPA is more accurate
(apparently exact), while the more sophisticated treatment introduces
errors.

The resolution of this paradox can be seen most easily in Eq. (1), 
the RPA-type equation for the susceptibility.
Insertion of $f\x$ alone (linear in $\lambda$) into these equations will lead to
all powers of $\lambda$ being present in the solution, since it is
a self-consistent integral equation.  This is most easily seen by iterating
the equations.  A simple way to recover the exact first-order GL result is
by solving the equations with $\lambda f\x$, and making $\lambda$ very small,
to find the linear contribution to the change in excitation energies.
Insertion of $\lambda=1$ into this result will yield the exact GL result.

A far simpler method is to use the single-pole approximation.
To see why this works, consider Eq. (2), our matrix whose eigenvalues
are at the excitation energies.  Since $f\Hxc$ is at least first-order
in $\lambda$, all off-diagonal contributions are of O($\lambda$).  Thus
to lowest order, the diagonal dominates, and the off-diagonal
corrections contribute $O(\lambda^2)$ corrections.  Retaining only
diagonal contributions, i.e., the single-pole approximation, yields the
exact result to first order in $\lambda$.
A detailed functional derivation of this result has recently been given by Gonze
and Scheffler\cite{GS99}.

\subsection{A new hybrid functional for the kernel}

\begin{figure}[t]
\vskip -2 cm
\unitlength1cm
\begin{picture}(10.5,12.5)
\put(-7,7)
{\makebox(14.5,8.5){
\includegraphics{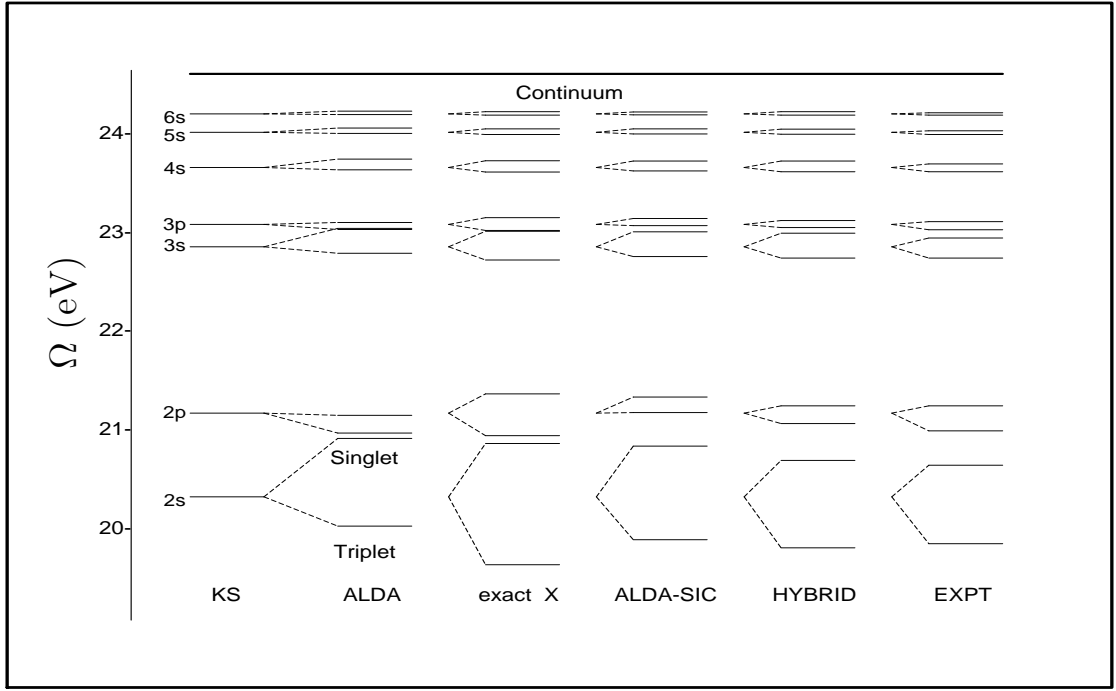}
}}
\setbox6=\hbox{\Large $\Omega$ (eV)}
\put(1.8,6.5) {\makebox(0,0){\rotl 6}}
\end{picture}
\vskip -1 cm
\caption{\sl Excitation energies from the
ground-state of He, including the orbital energies
of the exact Kohn-Sham potential, the
time-dependent
spin-split correction calculated
within four different functional approximations,
and experimental numbers
\protect
\cite{PGG98}.}
\label{Heexcit}
\end{figure}

To illustrate the importance of understanding the origin of errors
in density functionals,
we use the insight gained within the SPA in the previous sections
to construct a hybrid functional for $f\xc$.  We then apply this both
within SPA and in the full calculation.    Our original idea, as mentioned
in section II, was to use a simple self-interaction correction to produce
a better approximation than either exact exchange or ALDA, but the SIC
results
\newpage ~\vskip 9cm\noindent
of Tables I and II show this does not happen.

Consider Fig. 1, which illustrates the positions of energy levels
in the different schemes for the He atom.
Our first step is to consider the mean energies.  As pointed out in
subsection A, these are determined by the parallel contributions to
$f\xc$ and the Hartree term.
We have already seen how these two terms cancel exactly at the exchange
level, so that the mean energy is very good in such a calculation.
In ALDA,
the cancellation (exact for
He) of the exchange contributions is lost. This can be seen
in the large shifts in the ALDA mean energies in Table II and in Fig. 1.
Furthermore, the
remaining small parallel-spin correlation contribution can be expected
to be grossly overestimated by ALDA, since LDA ground-state correlation energies
are usually too large by a factor of 2 or 3.
Unfortunately, even SIC-ALDA is not exact for two-electron exchange, and
it also suffers from a poor mean energy.
Thus we recommend using only
exact exchange for the parallel-spin contribution.

On the other hand, the splitting is determined solely by anti-parallel
correlation contributions, and the direct term.
Thus, an exact exchange treatment
misses entirely the significant anti-parallel correlation contribution.
This error is highlighted by the far too large splittings in the
exchange results in Tables I and II, implying significant cancellation
between the direct and antiparallel correlation contributions.
So here we advocate use of ALDA.  
Since the splitting depends on anti-parallel contributions
to $f\xc$, but the SIC correction only applies to one
spin at a time, 
SIC-ALDA has exactly the
same splittings.  Our recommended hybrid is therefore
\ben
f\xc^{\uparrow\uparrow} = f\x^{\uparrow\uparrow},
~~~~~~f\xc^{\uparrow\dn} = f\xc^{\uparrow\downarrow\ {\rm ALDA}}.
\label{fxchyb}
\een
Results shown in Tables I and II indicate that this hybrid decreases almost
all errors over either exact exchange or ALDA.   On average, the decrease
is by about a factor of 3 for the He atom, but much less for Be (about 40\%).
Notable exceptions are the triplet transitions to $p$ states in He and
in Be, where the error is increased.

\section{Conclusions}

We have shown how the results of TDDFT with approximate
exchange-correlation  kernel functionals may be understood in terms
the well-known behavior of the ground-state functionals from which they
are derived.  We have shown in a simple case how a more accurate
functional may be constructed from this insight.  We  regard this as
an initial step toward an accurate approximation for $f\xc$.
Another obvious analytic tool would be the direct adiabatic
decomposition of $f\xc$ in terms of $\lambda$, which has proved so
successful in understanding the hybrids commonly used in 
ground-state calculations\cite{BEP97,PEB96}.

\acknowledgments
We gratefully acknowledge support from a NATO collaborative research
grant.  K.B. also acknowledges support of the Petroleum Research Fund
and of NSF grant no. CHE-9875091.  M.P. and E.K.U.G. also acknowledge
partial support of the DFG.

\end{document}